\documentclass[11pt]{article}
\usepackage{moriond}
\usepackage{graphicx}

\def\Journal#1#2#3#4{{#1} {\bf #2}, #3 (#4)}

\def\PRL{\em Phys. Rev. Lett.}
\def\PRB{{\em Phys. Rev.} B}
\def\JETPL{{\em JETP Lett.}}

\def\ab{\bar{\alpha}}
\def\be{\begin{equation}}
\def\ee{\end{equation}}
\def\bea{\begin{eqnarray}}
\def\eea{\end{eqnarray}}

\newcommand{\EXP}[1]{{\mbox{\large e}}^{#1}}         
\newcommand{\diagram}[3]{\raisebox{#3}{\includegraphics[scale=#2]{#1}}}
\def\I{{\rm i}}                  
\def\D{{\rm d}}                  
\def\ab{{\alpha\beta}}
\newcommand{\smean}[1]{\langle #1 \rangle} 
\newcommand{\APPROX}[1]{                
   {{\raisebox{-.3cm}{$\textstyle\simeq$}} \atop {\scriptstyle{#1}}}}

\begin{document}
\title{How to increase a transmission with weak localization~? 
       A geometrical effect}

\author{\underline{Christophe Texier}$^{1,2}$, Gilles Montambaux$^2$}

\address{$^1$Laboratoire de Physique Th\'eorique et Mod\`eles Statistiques.
Universit\'e Paris-Sud, B\^at. 100.}

\address{\vspace{-0.25cm}
$^2$Laboratoire de Physique des Solides.
Universit\'e Paris-Sud, B\^at. 510, F-91405 Orsay Cedex, France.}

\maketitle
\abstracts{
We study the quantum transport in multiterminal networks of 
quasi-one-dimensional diffusive wires.
When calculating the weak localization correction to the conductances, 
we show that the Cooperon must be properly weighted over each wire.
This can even change the sign of the weak localization correction in 
certain geometries.
}

\section{Introduction}

The classical conductance of a wire of length $L$ and section $s$ made of 
weakly disordered metal is given by the Ohm's law $G_{\rm D}=\sigma_0s/L$,
where $\sigma_0$ is the Drude conductivity.
Quantum interferences are responsible for a small {\it negative} contribution,
in the absence of spin-orbit coupling, that shows up at sufficiently low 
temperature, the so-called weak localization (WL) correction.
This effect, which is destroyed by a strong magnetic field, is particularly 
interesting from the experimental point of view because it provides an 
efficient way to extract a phase coherence length $L_\varphi$ in a disordered 
metal, by studying its magnetoconductance.
In the middle of the 80's, the progresses in nanolithography allowed to
realize not only wires of mesoscopic sizes ($\mu$m) but also networks
of wires, whose more complicate topologies make them particularly suitable 
to study interference effects. The first experiments were realized on large
honeycomb metallic lattices \cite{PanChaRamGan84} 
and showed the AAS oscillations predicted in \cite{AltAroSpi81}. 
Many other experiments have been realized until then on necklace of loops, 
ladders, square lattices,...
The first theoretical description of WL in networks was provided by 
Dou{\c c}ot \& Rammal (DR) \cite{DouRam85}
and was successfully used 
to describe the experiments of Pannetier {\it et al} \cite{PanChaRamGan84}. 
The starting point of this approach is a uniform integration of the Cooperon 
$\Delta\sigma(\vec r)=-\frac{e^2}{\pi}P_c(\vec r,\vec r)$ over the 
system \cite{footnote1}.
$P_c(r,r)$ is  the return probability of a diffusion problem.
However, it is meaningful to define a {\it local} conductivity 
($\sigma=\frac1{\rm Vol}\int\D\vec r\D\vec r\,'\,\sigma(\vec r,\vec r\,')$)
only if the sys-

\noindent
\begin{minipage}[b]{9cm}
  tem is regular. If instead one considers a
  network of arbitrary topology, like on figure 1,
  its tranport properties are conveniently described, in the 
  Landauer-B\"uttiker formalism, by transmission probabilities 
  $T_{\alpha'\beta'}$ to go from a contact $\beta'$ to a contact $\alpha'$, 
  which are related to the {\it nonlocal} conductivity.
  The transmissions averaged over the disorder are written
  $\smean{T_{\alpha'\beta'}}=T^{\rm cl}_{\alpha'\beta'}
  +\Delta T_{\alpha'\beta'}+\cdots$
  We have demonstrated in \cite{TexMon04} that the Cooperon must be properly 
  weighted when integrated over the wires $(\mu\nu)$ of the network. These 
  weights depend on the topology and the way the network is connected to 
  reservoirs. 
  Moreover, we showed
\end{minipage}
\hspace{0.5cm}
\begin{minipage}[b]{6.25cm}
  \hspace{-0.1cm}\includegraphics[scale=0.9]{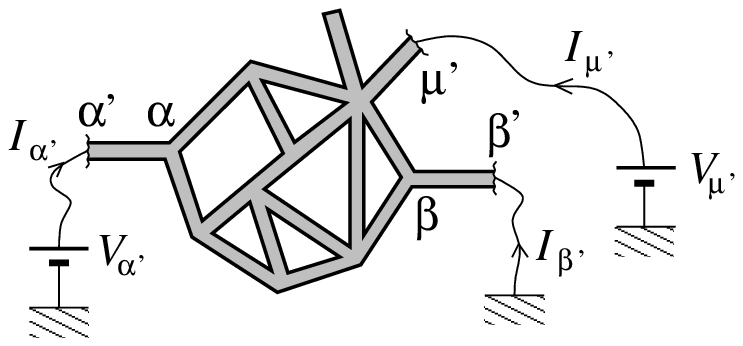}
  {\small
  Figure 1: A network of diffusive wires. The network is connected at 
  the vertices $\alpha'$, $\beta'$ and $\mu'$ to external reservoirs 
  (wavy lines) through which some current is injected in the network. 
  }
\end{minipage}
\setcounter{figure}{1}

\vspace{-0.4cm}

\noindent
that the geometrical weight to attribuate to a wire $(\mu\nu)$
is given by the derivative of the classical transmission
$T^{\rm cl}_{\alpha'\beta'}$ with respect to the length $l_{\mu\nu}$~:
\be\label{RES3}
\Delta T_{\alpha'\beta'} =
\frac{2}{\alpha_d N_c\ell_e} \sum_{(\mu\nu)} 
\frac{\partial T^{\rm cl}_{\alpha'\beta'}}{\partial\,l_{\mu\nu}}
\int_{(\mu\nu)}\D x\, P_c(x,x)
\:.\ee
$N_c$ is the number of conducting channels per wire, $\ell_e$ the elastic 
mean free path and $\alpha_d$ a numerical constant depending on the dimension
($\alpha_1=2$, $\alpha_2=\pi/2$ and $\alpha_3=4/3$). 
For regular networks the uniform integration of $P_c$ is justified, however
the weights cannot be neglected in general~:
In multiterminal geometries they can even 
{\it change the sign of the WL correction}.

\section{Transport in weakly disordered metals}

The classical transport is described by two terms~: 
the Drude contribution (short range) and the contribution from the 
diffuson (ladder diagrams) which is long range.
In the diffusion approximation, we have \cite{KanSerLee88}~:

\vspace{-0.6cm}

\bea\label{classcon}
\smean{\sigma_{ij}(\vec r,\vec r\,')}_{\rm class}
= \diagram{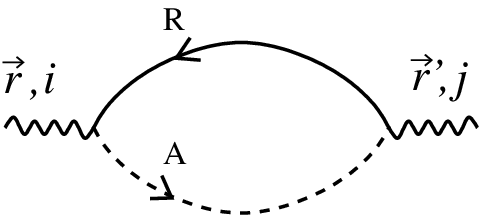}{0.5}{-0.4cm} 
+ \diagram{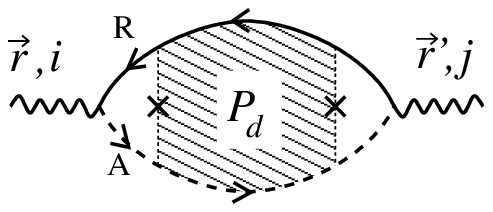}{0.5}{-0.6cm}
=\sigma_{\rm 0}\,
\left[
  \delta_{ij}\:{\delta}(\vec r-\vec r\,')
 -\nabla_i\nabla'\!\!_j P_d(\vec r,\vec r\,')
\right]
\:,\eea

\vspace{-0.3cm}

\noindent
where the diffuson $P_d$ is solution of the equation
$-\Delta P_d(\vec r,\vec r\,')=\delta(\vec r-\vec r\,')$.
We introduce the notation
$\smean{\sigma_{ij}(\vec r,\vec r\,')}_{\rm class}
=\sigma_{\rm 0}\: \phi_{ij}(\vec r,\vec r\,')$.
An important requirement of a transport theory is to satisfy current
conservation $\nabla_i\sigma_{ij}(\vec r,\vec r\,')=0$, what the classical
conductivity (\ref{classcon}) does.

The WL correction is given by the maximally crossed diagrams 
\cite{GorLarKhm79,AltKhmLarLee80}~:
%
%
\bea\label{GLK}
\smean{\sigma_{ij}(\vec r,\vec r\,')}_{\rm cooperon}
 = \diagram{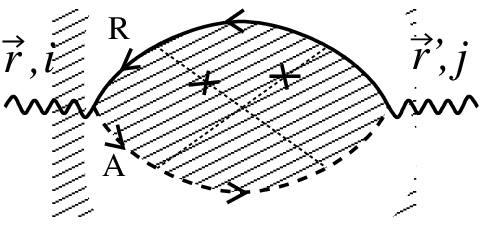}{0.5}{-0.6cm} 
= - \frac{e^2}{\pi}\,\delta_{ij}\,\delta(\vec r-\vec r\,')\,
P_c(\vec r,\vec r)
\:,\eea
%
%
\noindent
where the Cooperon is solution of
$
[\frac{1}{L_\varphi^2}-(\vec\nabla-2\I e\vec A)^2]
P_c(\vec r,\vec r\,') 
= \delta(\vec r-\vec r\,')
$.
It is clear that the additional contribution of the Cooperon 
(\ref{GLK}) does not respect current conservation. 
In the same way that the classical conductivity is built of short range 
(Drude) and long range (diffuson) contributions, the WL 
correction contains long range terms additionally to the short range 
contribution (\ref{GLK}). 
A convenient method to construct the current conserving quantity was
proposed in \cite{KanSerLee88}, that avoids appearance of divergencies 
in the long range contribution. It leads to \cite{TexMon04}
\bea\label{KSL}
\smean{\Delta\sigma_{ij}(\vec r,\vec r\,')}
= \int\D\vec\rho\,\D\vec\rho\,'\,
\phi_{ii'}(\vec r,\vec\rho)\,\phi_{jj'}(\vec r\,',\vec\rho\,')\,
\smean{\sigma_{i'j'}(\vec \rho,\vec \rho\,')}_{\rm cooperon}
\:.\eea

\section{Networks}

We now consider specifically the case of networks such as the one on 
figure 1. 
The transmission $T_{\alpha'\beta'}$ is related to the nonlocal conductivity 
with the two arguments at contacts $\alpha'$ and $\beta'$
(see details in \cite{TexMon04,TexMonAkk04}).
For quasi one-dimensional (1d) wires, we obtain the expressions
\bea\label{wlg}
T^{\rm cl}_{\alpha'\beta'}=\frac{\alpha_d N_c}{\ell_e}
P_d(\underline{\alpha}',\underline{\beta}')
\hspace{0.25cm}\mbox{ and }\hspace{0.25cm}
\Delta{T}_{\alpha'\beta'}=\frac2{\ell_e^2}
\int_{\rm Network}\hspace{-0.25cm}\D x\, 
\frac{\D}{\D x}P_d(\underline{\alpha}',x)\:
P_c(x,x)\:\frac{\D}{\D x}P_d(x,\underline{\beta}')
\:,\eea
that involve the 1d diffuson and cooperon.
The notation $P_d(\underline{\alpha}',x)$ means that $P_d$ is 
taken at a distance $\ell_e$ of the vertex $\alpha'$
(the question of boundary conditions is discussed in detail in
\cite{TexMon04,TexMonAkk04}).
Since $P_d(\underline{\alpha}',x)$ is a linear function of $x$, the two 
diffusons in $\Delta{T}_{\alpha'\beta'}$ bring a constant that depends on 
the wire to which $x$ belongs. 
In other terms, when integrated over a wire, $P_c$ must be weighted by a 
coefficient depending on the wire, shown in \cite{TexMon04} to be the 
one given in (\ref{RES3}).

The next step is to construct explicitely $P_{d,c}$ in the network.
To describe its topology we introduce the adjacency matrix 
$a_{\alpha\beta}$~: we have
$a_{\alpha\beta}=1$ if the vertices $\alpha$ and $\beta$ are connected by 
a wire and $a_{\alpha\beta}=0$ otherwise.
The parameters $\lambda_\alpha$ describe how the network is connected~:
$\lambda_\alpha=0$ for an internal vertex and $\lambda_{\alpha'}=\infty$
at the vertices connected to external reservoirs (the connected vertices
are primed). 
The magnetic flux along the wire $(\ab)$ is denoted $\theta_\ab$.
The solution for $P_c$ involves the matrix 
\cite{DouRam85,PasMon99,AkkComDesMonTex00,TexMonAkk04}~:
%
%
\bea
{\cal M}_\ab =
\delta_\ab
\left(\lambda_\alpha + \sqrt\gamma\sum_\mu a_{\alpha\mu}
      \coth(\sqrt{\gamma}l_{\alpha\mu})\right)
-a_\ab\frac{\sqrt\gamma\:\EXP{-\I\theta_\ab}}{\sinh(\sqrt{\gamma}l_\ab)}
\:.\eea
%
%
The diffuson is expressed in terms of the same matrix with $\gamma=0$ and 
no magnetic flux $\theta_\ab=0$~:
%
%
\be
\left({\cal M}_0\right)_\ab = 
\delta_\ab
\left(\lambda_\alpha + \sum_\mu a_{\alpha\mu}\frac1{l_{\alpha\mu}}\right)
-a_\ab\frac{1}{l_\ab}
\:.\ee
%
%
This matrix encodes the information about the classical conductances 
$\alpha_d N_c\ell_e/l_{\mu\nu}$ of each wire $(\mu\nu)$.
Note that $\lambda_{\alpha'}=\infty$ (at a reservoir)
implies that $({\cal M}^{-1})_{\mu\alpha'}=0$, $\forall$ $\mu$.
The classical conductance is given by
\be\label{RES1}
T^{\rm cl}_{\alpha'\beta'} =
\frac{\alpha_d N_c\ell_e}{l_{\alpha\alpha'}l_{\beta\beta'}} 
\left({\cal M}_0^{-1}\right)_\ab
\:.\ee
This result is only valid for $\alpha'\neq\beta'$ (see 
\cite{HasStoBar94,TexMonAkk04}). 
It coincides with the one obtained for 
a network of classical resistances, as it should. 
The integral of the Cooperon in (\ref{RES3}) is a nonlocal quantity that 
carries information on the whole structure of the network through the 
matrix ${\cal M}^{-1}$~:
\bea\label{RES2bis}
\int_{(\mu\nu)}\D x\,P_c(x,x)
=\frac1{2\sqrt\gamma}\bigg\{
\left[ 
\left({\cal M}^{-1}\right)_{\mu\mu}+
\left({\cal M}^{-1}\right)_{\nu\nu}
\right]
\left(
  \coth\sqrt\gamma l_{\mu\nu} 
- \frac{\sqrt\gamma l_{\mu\nu}}{\sinh^2\sqrt\gamma l_{\mu\nu}}
\right)\nonumber\\
+
\left[ 
 \left({\cal M}^{-1}\right)_{\mu\nu}\EXP{\I\theta_{\mu\nu}}
+\left({\cal M}^{-1}\right)_{\nu\mu}\EXP{\I\theta_{\nu\mu}}
+\frac{\sinh\sqrt\gamma l_{\mu\nu}}{\sqrt\gamma}
\right]
\frac{-1+  \sqrt\gamma l_{\mu\nu} \coth\sqrt\gamma l_{\mu\nu} }
     {\sinh\sqrt\gamma l_{\mu\nu}} 
\bigg\}
\:.\eea
Both the classical transmissions and their WL corrections can be computed 
by algebraic calculations only from the two matrices.

\section{Importance of the weights on two examples of networks}

We now study two examples of networks where the existence of the weights
cannot be neglected.

\noindent
\begin{minipage}[b]{7.5cm}
\diagram{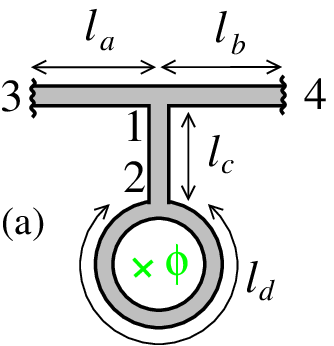}{0.9}{2cm}
\diagram{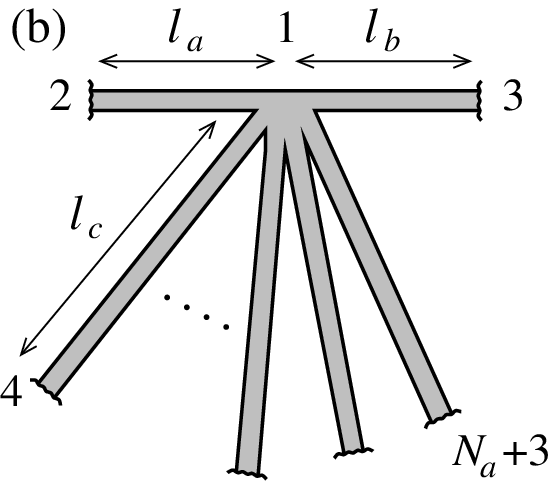}{0.7}{2cm}

\vspace{-2cm}

{\small Figure 2~:
         Two examples of mesoscopic devices. }

\vspace{0.15cm}
\end{minipage}
\hspace{0.5cm}
\begin{minipage}[b]{7.75cm}

\vspace{0.25cm}

{\it A ring}.--
The first example is the loop of figure 2a. 
This example 
emphasizes the nonlocality of the WL. The classical transmission 
$T^{\rm cl}_{34} = \frac{\alpha_d N_c\ell_e}{l_a+l_b}$  is independent
of the presence of the arm and the ring. 
Then their weights are 0 and this part of the network 
does not contribute to the WL correction (\ref{RES3}). 
However, since the cooperon $P_c(x,x)$ is {\it nonlocal},
it feels the presence of the loop, even for $x$ in the wires 
\end{minipage}

\noindent
$(3\!\leftrightarrow\!1)$ and $(1\!\leftrightarrow\!4)$.
The harmonics of the AAS oscillations,
$
\Delta \hat T_{34}^{(n)}
=\int_0^{\phi_0/2}\hspace{-0.1cm}
\frac{\D\phi}{\phi_0/2}\Delta T_{34}(\phi)\:
\EXP{-4\I\pi n{\phi}/{\phi_0}}
$,
are found to be, in the limit $l_a,l_b\gg L_\varphi$~:
\bea\label{harmT34a}
\Delta \hat T_{34}^{(n)}
&\simeq&
-\left(\frac{L_\varphi}{l_a+l_b}\right)^2\left(\frac23\right)^{n+3}
\EXP{-{2l_c}/{L_\varphi}-n\,{l_d}/{L_\varphi}}
\hspace{0.5cm}\mbox{for }l_c,l_d\gg L_\varphi
\\\label{harmT34b}
&\simeq&
-\left(\frac{L_\varphi}{l_a+l_b}\right)^2\sqrt{\frac{l_d}{2L_\varphi}}
\ \EXP{-n\sqrt{{2l_d}/{L_\varphi}}}
\hspace{1.6cm}\mbox{for }l_c,l_d\ll L_\varphi
\eea
Note that the naive uniform integration of the cooperon over the network
(DR\&PM) strongly overestimates the amplitude of the AAS oscillations 
(figure 3) 
\cite{TexMonAkk04}. This is related to the fact that the uniform 
integration misses the $\EXP{-2l_c/L_\varphi}$ factor brought by the 
Cooperon exploring the ring starting from the wire.
The decrease of the WL at high field (inset) is due to the 
contribution of the flux to the effective phase coherence 
length \cite{AltAro81} $L_\varphi(\phi)$.
The behaviour $\EXP{-n\,l_d/L_\varphi}$ in (\ref{harmT34a}) comes from the
normal diffusion in a ring ($n_t$, the typical number of turns after a 
time $t$ behaves like $n_t\sim t^{1/2}$).
On the other hand, for a very coherent ring ($l_d\ll L_\varphi$) connected 
to long wires ($l_a,l_b\gg L_\varphi$), the change of behaviour in 
(\ref{harmT34b}) originates from a subdiffusive motion ($n_t\sim t^{1/4}$)
due to the exploration of the long wires ($\gg l_d$)~:
each time the diffusive trajectory encircles the loop, the diffusion in 
the wires increases the effective perimeter, that now reads 
$l_{\rm eff}=l_d\sqrt{2L_\varphi/l_d}$. 
The same phenomenon occurs in the study of the transport through a ring.
This illustrates that the connecting wires must be properly taken into
account.

\noindent
\begin{minipage}[b]{8cm}
\diagram{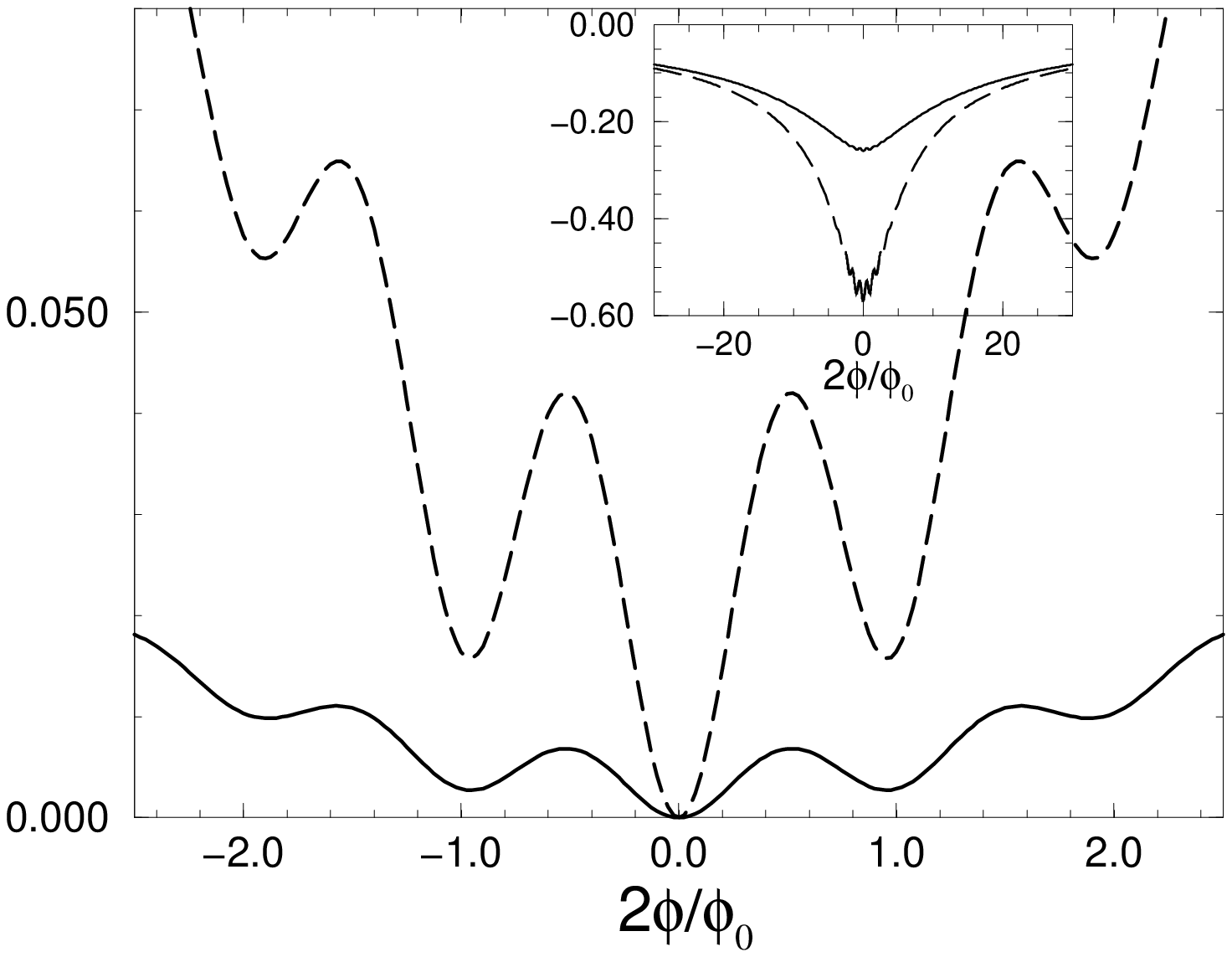}{0.5}{0cm}
\end{minipage}
\begin{minipage}[b]{7cm}
{\small Figure 3~:\\
         Magnetoconductance for the ring of figure 2a.\\
         Dashed line~: $\Delta\sigma/\sigma_0$ for a uniform
         integration of $P_c$.
         Continuous line~: $\Delta T_{34}/T^{\rm cl}_{34}$ 
         given by (\ref{RES3}).
         The curves have been shifted so that they coincide at $\phi=0$.\\
         {\it Inset~:}
         Same curves (without shift) for a higher window of 
         flux $\phi$.
         The parameters are 
         $l_a=l_b=1\:\mu$m, $l_c=0.05\:\mu$m and $l_d=5\:\mu$m.
         $W=0.19\:\mu$m, $L_\varphi(\phi=0)=1.7\:\mu$m. }

\vspace{1.3cm}
\end{minipage}


\noindent{\it Multiterminal geometry}.--
The origin of the negative WL correction $\Delta T_{\alpha'\beta'}$
lies in the negative weights, however for a multiterminal geometry some 
weights can be {\it positive}, like the weight(s)
$\partial{T^{\rm cl}_{\alpha'\beta'}}/\partial{l_{\mu\mu'}}$ of the wire(s)
connected to other terminal(s) ({\it e.g.} $\mu'$ on figure 1). 
As a first example, we consider a wire on which is plugged one long arm of
length $l_c$ connected to a third reservoir (figure 2b 
for $N_a=1$).
We focus ourselves on the fully coherent limit $L_\varphi=\infty$.
The classical (Drude) conductance of this 3-terminal network is~:
$T^{\rm cl}_{23}=\frac{\alpha_d N_c\ell_e\,l_c}{l_al_b+l_bl_c+l_cl_a}$.
Then ${\partial T^{\rm cl}_{23}}/{\partial l_c}>0$.
The wire [$(2\!\leftrightarrow\!1)+(1\!\leftrightarrow\!3)$] gives a negative 
contribution to the WL correction whereas the arm 
$(1\!\leftrightarrow\!4)$ gives a {\it positive}
one. Introducing $l_{a/\!/b/\!/c}^{-1}=l_a^{-1}+l_b^{-1}+l_c^{-1}$, we find
\be\label{Zewire}
\Delta T_{23}
=\frac13
\left(
  -1 + \frac{l_{a/\!/b/\!/c}}{l_c} + \frac{l_{a/\!/b/\!/c}^2}{l_al_b} 
\right)
\APPROX{l_{a/\!/b}\ll l_c}\frac13\left(-1+\frac{l_{a/\!/b}}{l_a+l_b}\right)
\:.\ee
We now consider the case of $N_a$ long arms plugged in the middle of the wire
($l_a=l_b$)  like on figure 2b, 
to maximize their effect \cite{TexMonAkk04}. We obtain~:
\be\label{powl}
\Delta T_{23}\simeq\frac13\left(-1 + \frac{N_a}{4}\right)
\:,\ee 
a result valid for $l_a\ll l_c\ll L_\varphi$. 
We can now obtain a positive WL correction for $N_a>4$. 
This effect is purely geometrical.
Note that in the limit $l_c\gg L_\varphi$ the positive 
contribution vanishes.
The positive contribution of the arm in (\ref{Zewire}) can only be 
observed in a 3-terminal measurement, when the arm is a current sink. 
If the arm is a voltage probe, the positive contribution vanishes.


\begin{thebibliography}{10}
\bibitem{footnote1}
  This approach has been made more efficient by Pascaud \& Montambaux (PM)
   in \cite{PasMon99}.
\end{thebibliography}

\end{document}